\documentclass[iop]{emulateapj}
\usepackage{natbib}
\usepackage{longtable} 
\usepackage{amsmath} 
\usepackage{color} 

\def\dt{\Delta t}
\def\cov{S}
\def\sf{SF}
\def\psd{P}
\def\pe{powered--exponential}
\def\ma{Mat\'ern}
\def\pa{Pareto--exponential}
\def\ke{Kepler--exponential}

\def\tcut{\tau_\mathrm{cut}}

\shorttitle{Quasar Variability Models} 
\shortauthors{Y.Zu, C.S. Kochanek, S. Koz\l{}owski, \& A. Udalski}
\begin{document}
\title{Is Quasar Optical Variability a Damped Random Walk?}
\author{Ying Zu\altaffilmark{1}, C.S. Kochanek\altaffilmark{1,2},
Szymon Koz\l{}owski\altaffilmark{3},
and Andrzej Udalski\altaffilmark{3}
}
\altaffiltext{1}{Department of Astronomy, The Ohio State University, 140 West
18th Avenue, Columbus, OH 43210, USA; yingzu@astronomy.ohio-state.edu}
\altaffiltext{2}{The Center for Cosmology and Astroparticle Physics, The Ohio
State University, 191 West Woodruff Avenue, Columbus, OH 43210, USA}
\altaffiltext{3}{Warsaw University Observatory, Al. Ujazdowskie 4, 00-478
Warszawa, Poland}
%
\begin{abstract} 
The damped random walk~(DRW) model is increasingly used to model the
variability in quasar optical light curves, but it is still uncertain whether
the DRW model provides an adequate description of quasar optical variability
across all time scales. Using a sample of OGLE quasar light curves, we consider
four modifications to the DRW model by introducing additional parameters into
the covariance function to search for deviations from the DRW model on both
short and long time scales.  We find good agreement with the DRW model on
time scales that are well sampled by the data~(from a month to a few years),
possibly with some intrinsic scatter in the additional parameters, but this
conclusion depends on the statistical test employed and is sensitive to
whether the estimates of the photometric errors are correct to within $\sim$$
10\%$. On very short time
scales~(below a few months), we see some evidence of the existence of a cutoff
 below which the correlation is stronger than the DRW model, echoing
the recent finding of~\cite{mushotzky2011} using quasar light curves from
{\it Kepler}. On very long time scales~($>$ a few years), the light curves
do not constrain models well, but are consistent with the DRW model.
\end{abstract}

\section{Introduction}
\label{sec:intro}

The optical variability of quasars has long been a proposed diagnostic
of the central engines of quasars~\citep[e.g.,][]{kawaguchi1998}. While
the physical origins of variability remain an open question~\citep[see,
e.g.,][]{dexter2011}, recently it has been proposed that the optical
variability of quasars is mathematically well described by a damped random
walk~\citep[DRW,][]{kelly2009}, which characterizes quasar light curves as a
stochastic process with an exponential covariance function $\cov(\Delta t)
= \sigma^2 \exp(-|\Delta t/\tau|)$, defined by an amplitude $\sigma$ and a
characteristic time scale $\tau$. \cite{kelly2009}, \cite{kozlowski2010} and
\cite{macleod2010} have shown that the DRW model provides a viable explanation
for the variability of individual quasars found in a heterogeneous sample,
OGLE~\citep[Optical Gravitational Lensing Experiment,][]{udalski1997} and SDSS
Stripe 82~\citep[S82,][]{sesar2007}, respectively, and that the parameters
$\tau$ and $\sigma$ are correlated with the physical properties of the
quasar~(rest wavelength, luminosity, and black hole mass). \cite{macleod2011-1}
also demonstrated that the {\it ensemble} structure functions of quasars
can be recovered by modeling each individual quasar with a DRW based on its
luminosity, rest wavelength and estimated black hole mass.

The DRW model is also a powerful tool. For example, it provides a well defined
approach to variability-selecting quasars~\citep{kozlowski2010,
macleod2011, butler2011}. It also provides a well-defined approach to
interpolating and modeling quasar light curves in reverberation mapping studies.
\cite{zu2011} show how it provides a general approach to measuring lags in one
or more emission lines or velocity bins of a single line. The model has been
used by~\cite{grier2012}, and a variant was used by~\cite{pancoast2011}.

However, it is still uncertain whether the DRW model, with only two parameters,
can fully characterize quasar optical variability across all timescales.  Early
studies~\citep[e.g.][]{collier2001} examined the power spectrum density~(PSD) of
a small number of individual quasars, finding evidence for the $f^{-2}$ power
law behavior of the DRW model on short time scales, and weaker evidence for a
flattening on long time scales. The first series of DRW
studies~\citep{kelly2009, kozlowski2010, macleod2010} essentially followed these
results and reconfirmed the average PSD structure, but did not explore any
possible deviations. We know from~\cite{mushotzky2011} that the DRW model
probably fails on very short time scales~($\lesssim$ week), having too much
power on short time scales compared to that measured by {\it Kepler}.
Mathematically speaking, the DRW model~(a.k.a. the Ornstein--Uhlenbeck process)
is unique~---~it is the only stationary Gaussian process~(GP) that is
``memoryless''~(Markov). \cite{kelly2009} introduced DRW model as a first order
continuous autoregressive (CAR(1))~process, and its discrete counterpart~(AR(1))
is simply an iteration $x_{i+1} = \alpha_\mathrm{AR}x_i + \epsilon_i$, where
$\epsilon$ is a Gaussian deviate.\footnote{\label{note:simu}In fact, a DRW light
curve is easily generated by using an iterative process with
$\alpha_\mathrm{AR}=e^{-|t_{i+1}-t_{i}|/\tau}$ and setting the variance of the
Gaussian deviate $\epsilon_i$ to be
$\sigma^2_\mathrm{AR}=\sigma^2(1-e^{-2|t_{i+1}-t_{i}|/\tau})$.} Physically, if
quasar variability is driven by multiple mechanisms, or multiple seeds of the
same mechanism with different relaxation time scales or amplitudes, there should
be deviations from the DRW model.

Traditional quasar variability studies often rely on examining the PSD~(or
the structure function as its real space equivalent) of the light curves.
PSD--based methods have two major drawbacks, one in measurement and one in
interpretation. In particular, the measurement of the PSD requires binning
the light curves, which are generally irregularly sampled for various
reasons~(weather, observing seasons, etc.). Information is always lost by
binning. More importantly, the uncertainties between the binned points are
very strongly correlated, making any statistical interpretation difficult
without correctly estimating the covariance matrix. While it is possible to
reconstruct the full covariance matrix of the empirical PSD using Monte Carlo
methods~\cite[e.g.,][]{uttley2002}, in this study we simply build a statistical
model of the light curve as a Gaussian process in the time domain, where
no discretization of data is involved, and it is easy to include
all the model/data covariances. This makes our approach more statistically
powerful than turning the light curve into an estimate of the PSD and then
fitting a model to the estimated PSD at the minor price of never producing
a visually appealing picture of the PSD (or structure function).

In this paper, we explore several alternative covariance functions with a third
parameter that allows for deviations from the DRW model and explore whether
they provide a better representation of quasar variability. This needs to be
done for individual quasars because we are now certain that the {\it ensemble}
statistics~(e.g., structure functions) of quasars are weighted averages of
individual quasars with different individual statistics~\citep{macleod2011}.  We
employ the OGLE light curves of quasars behind the Small and the Large
Magellanic Cloud (SMC and LMC). Thanks to the high-cadence and long-baseline of
OGLE, the light curves generally have $\sim 570$ photometric epochs over $\sim
7$ yrs, making them extremely well-suited for studies of quasar variability. We
describe this data set and our sample selection in Section~\ref{sec:data}. In
Section~\ref{sec:covfunc} we introduce four new covariance functions that we use
to test for deviations from the DRW model and describe the model fitting
procedures. We present our main results in Section~\ref{sec:result}.  We
conclude by discussing the physical implications in Section~\ref{sec:dis}.

\section{OGLE Quasar Light Curves}
\label{sec:data}

We use the light curves of quasars behind the LMC and SMC monitored by
OGLE~(\citealt{udalski1997, udalski2008}). Most of these were identified by
\cite{kozlowski2011, kozlowski2012} in part from candidates variability-selected
using the DRW model~\citep{kozlowski2010}. The photometric uncertainties of the
light curves were estimated by using Difference Image
Analysis~\citep[DIA,][]{wozniak2000}. The initial sample consists of $223$
I-band quasar light curves. There are typically $\sim 570$ epochs taken over
$\sim$ seven years on a two day cadence with six month seasonal gaps when OGLE instead
focuses on the Galactic bulge. For comparison, the S82 quasars have only $\sim
50$ epochs nominally over $\sim 10$ years, but there are really a few epochs in
1998, a gap until 2000 with 1--3 epochs per year coverage to 2005, and then
10--30 epochs per year from 2005 to 2008.  The OGLE quasars should allow us to
investigate quasar variability on both small and long timescales in more detail
than the SDSS S82 light curves.

Since we are examining the problem of additional parameters, we need higher
quality light curves than if we were only trying to estimate DRW parameters.
Starting from the $223$ light curves, we first selected sources with high S/N
for their variability, keeping sources where the rms magnitude variation
$\sigma_\mathrm{rms}$ was significantly larger than the mean photometric error
$\sigma_m$~($\sigma_\mathrm{rms}/\sigma_m > 2.0$). We also required $\sigma_m <
0.1\;\mathrm{mag}$. The typical photometric error for bright stars in OGLE is
$0.01\;\mathrm{mag}$, but the typical quasars are relatively fainter and so have
larger photometric uncertainties because there are only $\sim 3$ QSOs per
square degree at I$\leqslant$19 mag~\citep{richards2006}. This left us with 87
light curves in the sample. Next we required that the light curves yield a
well-defined DRW time-scale~$\tau$ by requiring that both $\ln
\mathcal{L}_{0}/\mathcal{L}_\mathrm{max} < -0.5$ and $\ln \mathcal{L}_{\infty} /
\mathcal{L}_\mathrm{max} < -0.5$, where $\mathcal{L}_{0}$,
$\mathcal{L}_{\infty}$ and $\mathcal{L}_\mathrm{max}$ are the likelihoods for
$\tau = 0$, $\tau = \infty$ and the best-fit $\tau$, respectively, where the
best--fit $\tau$ are in the range of $17 \lesssim \tau \lesssim 2700$ days.
This is to ensure that the light curves have some statistically detectable
characteristic timescale. The final sample has $55$ light curves of quasars with
redshifts between $0.15$ and $2.50$ and I-band apparent magnitudes between
$16.7$ and $19.4$ mag.

About $45\%$ of the objects are at redshifts where the I-band continuum
variabilities are contaminated by the lagged variabilities of broad emission
lines. For a broad band filter like I-band and assuming a typical quasar
spectrum~\citep{vanden_berk2001}, the line contributes only $\lesssim 6\%$
of the flux in the filter~(except for two objects at $z\sim 2.45$ where the
H$\alpha$ line contributes up to $14\%$ of the I-band flux). Since we are
probing the shape of the covariance functions instead of the variability
amplitudes, we do not expect this small level of contamination
to affect our results.  We tested for any effects by dividing our
sample into line-contaminated and line-free sub--samples, and found
negligible differences between the two sub--samples in the likelihood
analyses. Throughout the paper we fit logarithmic light curves~(i.e., in
magnitudes, as in~\citealt{kelly2009,kozlowski2010,macleod2010}), although we
used linear flux scales in~\cite{zu2011}. There is currently no theoretical
argument favoring fitting on one scale over another. Observationally, when
the variability amplitude of quasars is modest, we have found that fitting
on log and linear scales give almost identical results. Since the results
seem little affected by either broad line contamination or the scales used
in fitting, we will not discuss these issues further.

\section{Covariance Functions and Analysis Method}
\label{sec:covfunc}
 
\begin{figure}[t] 
\epsscale{1.0} 
\plotone{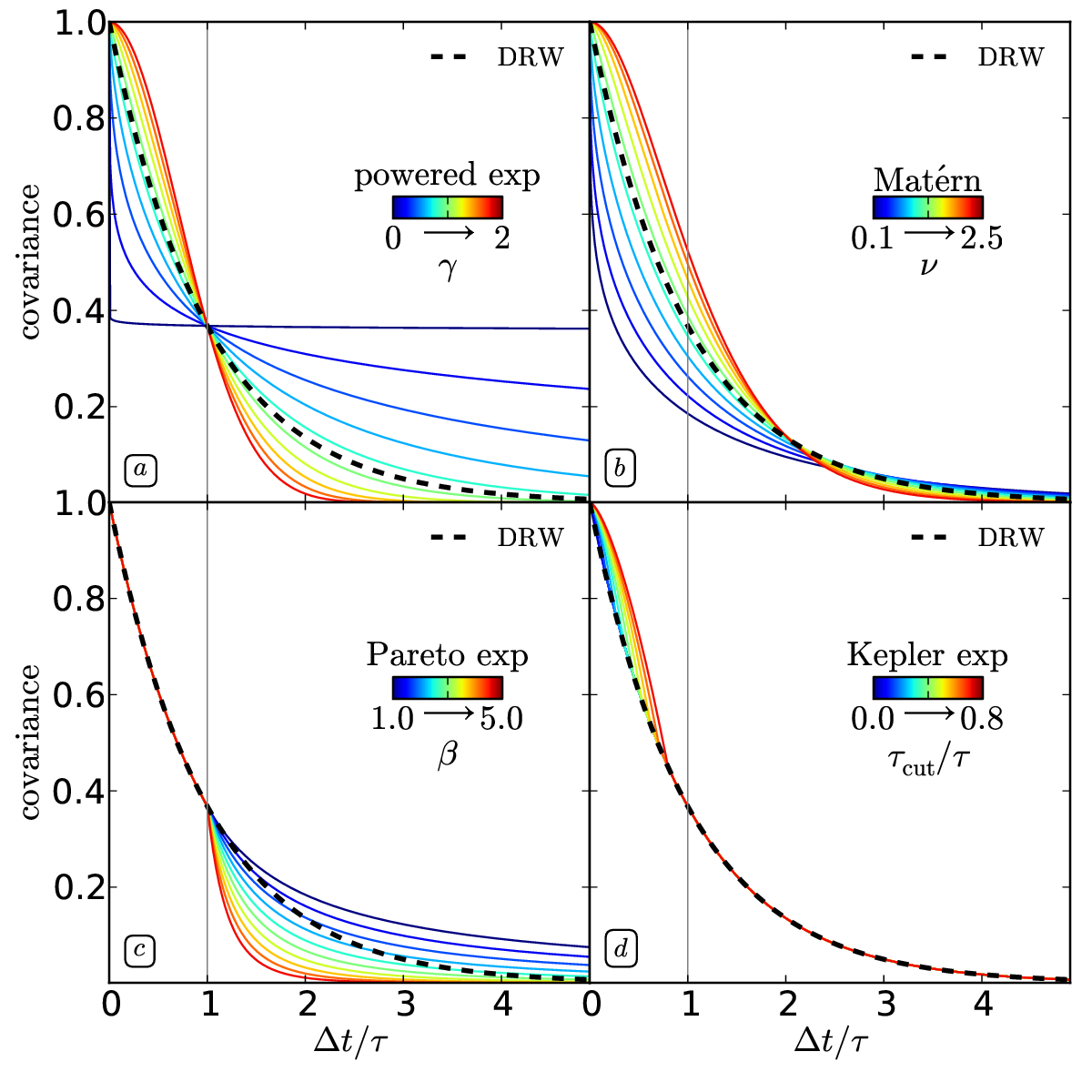} 
\caption{Comparison of the \pe~(panel $a$),
\ma~(panel $b$), \pa~(panel $c$), and \ke~(panel $d$) covariance functions~(thin
solid curves) with the DRW model~(thick dashed curves). The third parameter of
each model is given by the gray--scale colorbar in each panel. The models are
normalized to unity at zero lag.}
\label{fig:covdemo}
\end{figure}
 
We model each light curve as a Gaussian process,\footnote{In principle there
    could be deviations of light curves from Gaussian process, but the impact
    should be small given the $\chi^2$ distribution of the best--fit
    models~(see, e.g., Figure 1 of Zu et al. (2011)).} defined by a covariance
function $\cov(\dt) = \mathrm{cov}( m(t), m(t+\dt) )$ where $\dt$ is the time
separating two observations and we assume a stationary
process with a well-defined mean. The covariance function
is related to the structure function~(\sf) by
\begin{equation}
\sf(\dt) = \sqrt{2\sigma^2 - 2 \cov(\dt)}
\label{eqn:sf}
\end{equation}
and to the PSD by a Fourier Transform 
\begin{equation}
\psd(f) = \int\cov(\dt)e^{-2\pi\dt f i}d\dt.
\end{equation}
 
In particular, the DRW model has a covariance function of 
\begin{equation} 
\cov_\mathrm{DRW}(\dt) = \sigma^2
\exp\left({-\bigg|\frac{\dt}{\tau}}\bigg|\right)
\end{equation}
and a PSD of
\begin{equation} 
\psd(f) = \frac{4\sigma^2\tau}{1+(2\pi\tau f)^2}.
\label{eqn:psddrw}
\end{equation}
We do not repeat the SFs of the individual models because they are trivially
related to the covariance functions~(Equation~\ref{eqn:sf}). The DRW model can
be equivalently parameterized by $\tau$ and the asymptotic amplitude of the
\sf~\cite[$\sf_\infty=\sqrt{2}\sigma$,][]{macleod2010}, or $\tau$ and the slope
of the \sf~on short time
scales~\cite[$\hat{\sigma}=\sqrt{2\sigma^2/\tau}$,][]{kelly2009}.

\begin{figure}[t] 
\epsscale{1.00} 
\plotone{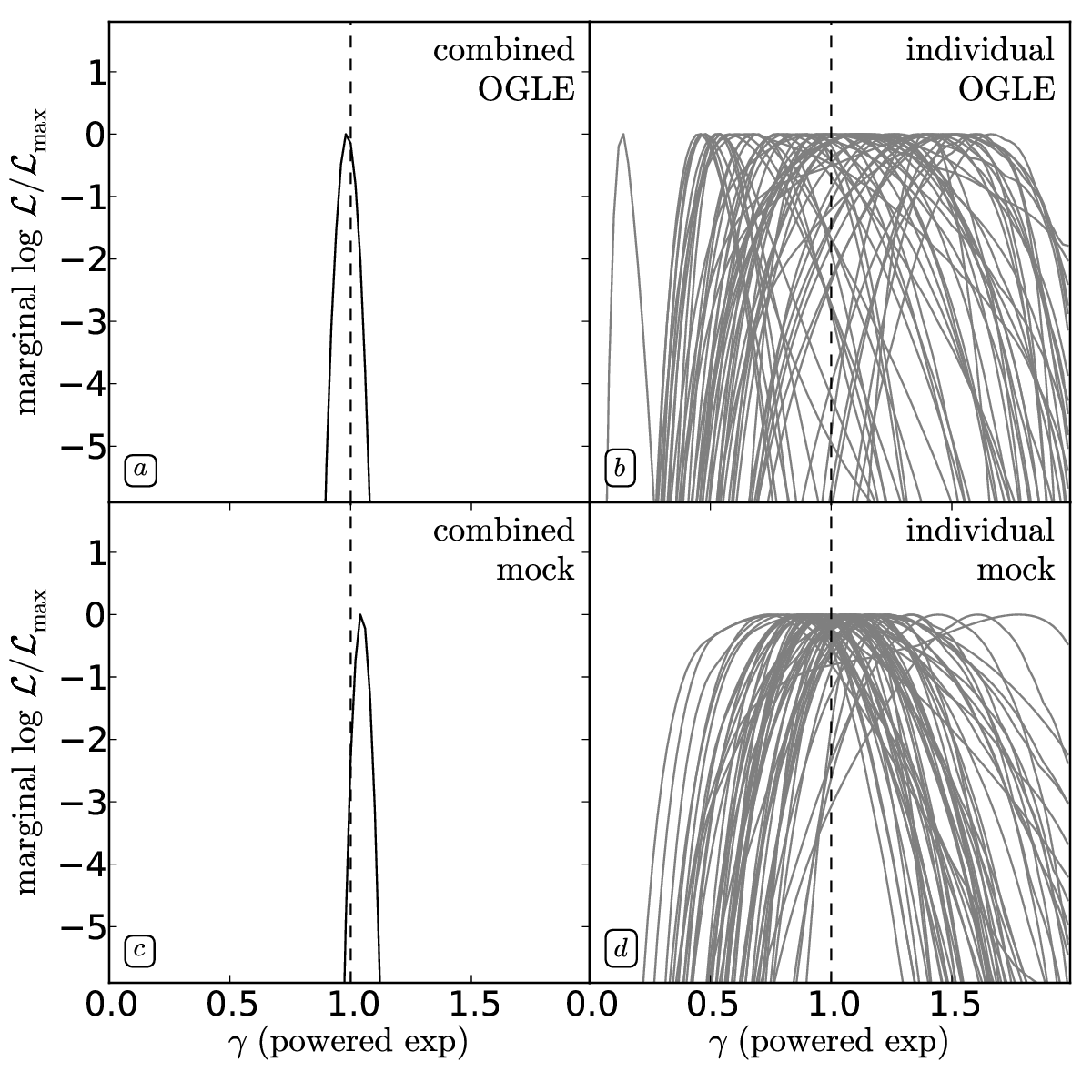}
\caption
{Combined~(left panels: a and c) and individual~(right panels: b and d) marginal
log-likelihoods for
$\gamma$ in the \pe~model for the data~(top panels: a and b) and the mock
data~(bottom
panels: c and d). The combined results are the sums of the log--likelihoods of
the individual quasars. For the DRW model $\gamma=1.0$, as indicated by the vertical
dashed line.}
\label{fig:pe}
\end{figure}

We consider the four modified
covariance functions shown in Figure~\ref{fig:covdemo}. A natural variant of the
DRW covariance function is the so-called ``\pe''~(PE) function~\citep{neal1997}
\begin{equation}
\cov_\mathrm{PE}(\dt) = \sigma^2
\exp\left({-\bigg|\frac{dt}{\tau}}\bigg|^\gamma\right)\quad\text{for $0<\gamma\le2$} ,
\end{equation}
where the special case $\gamma=1$ corresponds to the DRW model. The PSD for the
PE covariance can be expressed in a closed form only for $\gamma=1$~(DRW) and
$2$~(Gaussian), but for small time scales~($s \to \infty$), $\psd(s)\propto
s^{-(\gamma+1)}$~\citep{lim2003}.  The PE model modifies the DRW model on both
long and small scales~(panel $a$ of Figure~\ref{fig:covdemo}) and is the model
used in~\cite{pancoast2011}.

The ``\ma''~(MA) covariance function~\citep{matern1960} is defined by
\begin{equation} 
\cov_\mathrm{MA}(\dt) = \sigma^2
\frac{2^{1-\nu}}{\Gamma(\nu)}\left(\frac{\sqrt{2\nu}|\dt|}{\tau}\right)^\nu
K_\nu\left(\frac{\sqrt{2\nu}\dt}{\tau}\right)
\end{equation} 
where $\Gamma$ is the gamma function, $K_\nu$ is a modified Bessel function of
the second kind, and the new parameter, $\nu$ primarily adjusts the correlation
function on short time scales~($\nu>0$). The covariance function is exponential on long time scales
$\cov_\mathrm{MA}(\dt) = \sigma^2
\sqrt{\pi}\Gamma(\nu)^{-1}(\sqrt{\nu/2}|\dt|/\tau)^{\nu-1/2}\exp^{-\sqrt{2\nu}|\dt|/\tau}
$ as $\dt \to \infty$. The MA covariance function has a
PSD~\citep{abramowitz1965} of
\begin{equation}
\psd(f) =
\sqrt{\frac{2\pi}{\nu}}\frac{\tau\Gamma(\nu+1/2)}{\Gamma(\nu)}\left(1+\frac{4\pi^2\tau^2f^2}{2\nu}\right)^{-(\nu+1/2)}
\end{equation}
The special case $\nu=0.5$ gives the DRW covariance function~(panel $b$ of
Figure~\ref{fig:covdemo}).

\cite{mushotzky2011} found that the quasar light curves measured by {\it
Kepler} do deviate from the DRW model on very short time scales~($\sim$ days),
with steeper PSD power-law slopes of $-2.6$ to $-3.3$, rather than the $-2$ of
the DRW model~(Equation~\ref{eqn:psddrw}). Thus, there must be a $\tcut$ below which the fluctuation
amplitudes cut off more sharply than in the DRW model.  The OGLE light curves,
with a typical cadence of $\sim 2$ days, overlap the time scales of the {\it
Kepler} light curves~($\sim 6$ hr to $\sim 1$ month), so there is some prospect
of searching for a short time scale cutoff. To search for the cutoff within our
sample light curves, we define the \ke~(KE) covariance function by modifying
the DRW model on very short time scales to be
\begin{eqnarray}
\cov_\mathrm{KE}&&(\dt) = \nonumber\\
\sigma^2&&
\begin{cases}
1 -
(1-\exp\left(-|\frac{\tcut}{\tau}|\right))\left|\frac{\dt}{\tcut}\right|^{3/2} & \dt \le
\tcut \\
\exp\left({-|\frac{\dt}{\tau}}|\right)
& \dt > \tcut,
\end{cases}
\label{eqn:covke}
\end{eqnarray}
where $\tcut$ is the cutoff time scale below which the covariance function is
stronger than the DRW model, so that the PSD has a similarly steep slope as
found for the {\it Kepler} light curves. The model returns to the DRW model for
$\tcut=0$~(panel $d$ of Figure~\ref{fig:covdemo}), although we should not be
able to recognize values of $\tcut$ on scales smaller than a few times the
typical cadence~($\sim 10$ days).

\begin{figure}[t] 
\epsscale{1.00} 
\plotone{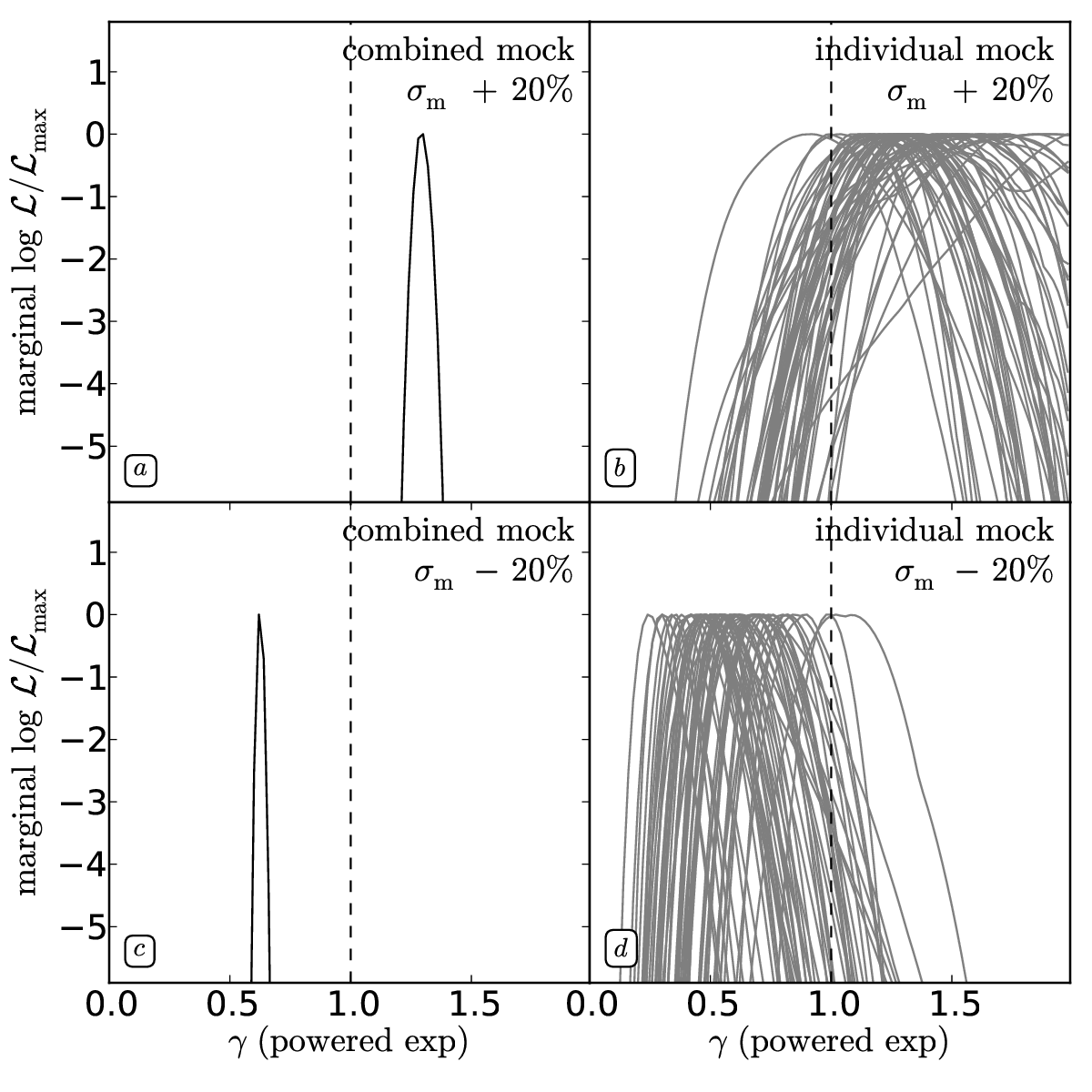}
\caption{The effects of incorrect error estimates for the \pe~model. The top~(bottom) panels
assume that the true photometric errors are over~(under) estimated by $20\%$. As
in Figure~\ref{fig:pe}, the left~(right) panels show the joint~(individual)
estimate of the new parameter $\gamma$.The effects of mis--estimating the
photometric errors are qualitatively similar for the other three models.}
\label{fig:errtest}
\end{figure}

In order to test for deviations from the DRW on long timescales, we combined
the DRW model on short timescales with the ``Pareto''
function~\citep{pareto1967} on long timescales~(the ``\pa'', PA), 
\begin{equation}
\cov_\mathrm{PA}(\dt) = \sigma^2
\begin{cases}
\exp({-\big|\frac{\dt}{\tau}}\big|) & \dt \le \tau \\
\frac{1}{e}\big|\frac{\dt}{\tau}\big|^{-\alpha} & \dt > \tau
\end{cases}
\end{equation}
where the power-law index $\alpha > 1$ controls the extent of the covariance
tails~(panel $c$ of Figure~\ref{fig:covdemo}). In this case,
no value of $\alpha$ exactly matches the DRW and placing the break at time scale
$\tau$ is reasonable but arbitrary.

Our goal is simply to estimate a value for the new parameters introduced in each
of these models using a Maximum Likelihood~(ML) approach. We briefly recap the
components of the likelihood function here and refer readers to~\cite{zu2011},
\cite{kozlowski2010} and ~\cite{rybicki1992} for details. We model each light
curve as
\begin{equation}
\mathbf{m}(t) = \mathbf{s}(t) + \mathbf{n} + L\mathbf{q},
\end{equation}
where $\mathbf{s}(t)$ is the underlying variability signal with covariance
matrix $S$,\footnote{The entries of $S_{ij}$ are simply the values of the
covariance function $S_{ij}=S(\dt_{ij})$, so we have used the same symbol for
both} $\mathbf{n}$ is the measurement uncertainty with covariance matrix $N$,
$L$ is a vector with all elements equal to one, and $\mathbf{q}$ is the light
curve mean.\footnote{The simultaneous fit of $\mathbf{q}$ is important
because any constant level of contamination (e.g., host galaxy light) is
removed by marginalizing over $\mathbf{q}$. Although we are fitting on log
scales, to first order the constant flux contribution can be ``subtracted'' in
magnitudes as well.} After optimizing the value of
$\mathbf{q}$, the likelihood of the model parameters is
\begin{eqnarray}
\mathcal{L}(\tau, \sigma,\text{$[\gamma, \nu, \tcut, \alpha]$}) =\hspace{.9cm}&&\nonumber\\
|C|^{-1/2}|L^T C^{-1}L|^{-1/2}&&\exp
\left(-\frac{\mathbf{m}^TC_{\perp}^{-1}\mathbf{m}}{2}\right)
\label{eqn:likelihood}
\end{eqnarray}
where $C$ is the overall data covariance $S+N$ and 
\begin{equation}
C_{\perp}^{-1} = C^{-1}-C^{-1}L(L^TC^{-1}L)^{-1} L^TC^{-1}.
\end{equation}
This approach needs no binning of the data and automatically includes all the
data and model uncertainty covariances, making it broadly superior~(in a
statistical sense) to standard uses of the PSD or structure function.

\begin{figure}[t] 
\epsscale{1.00} 
\plotone{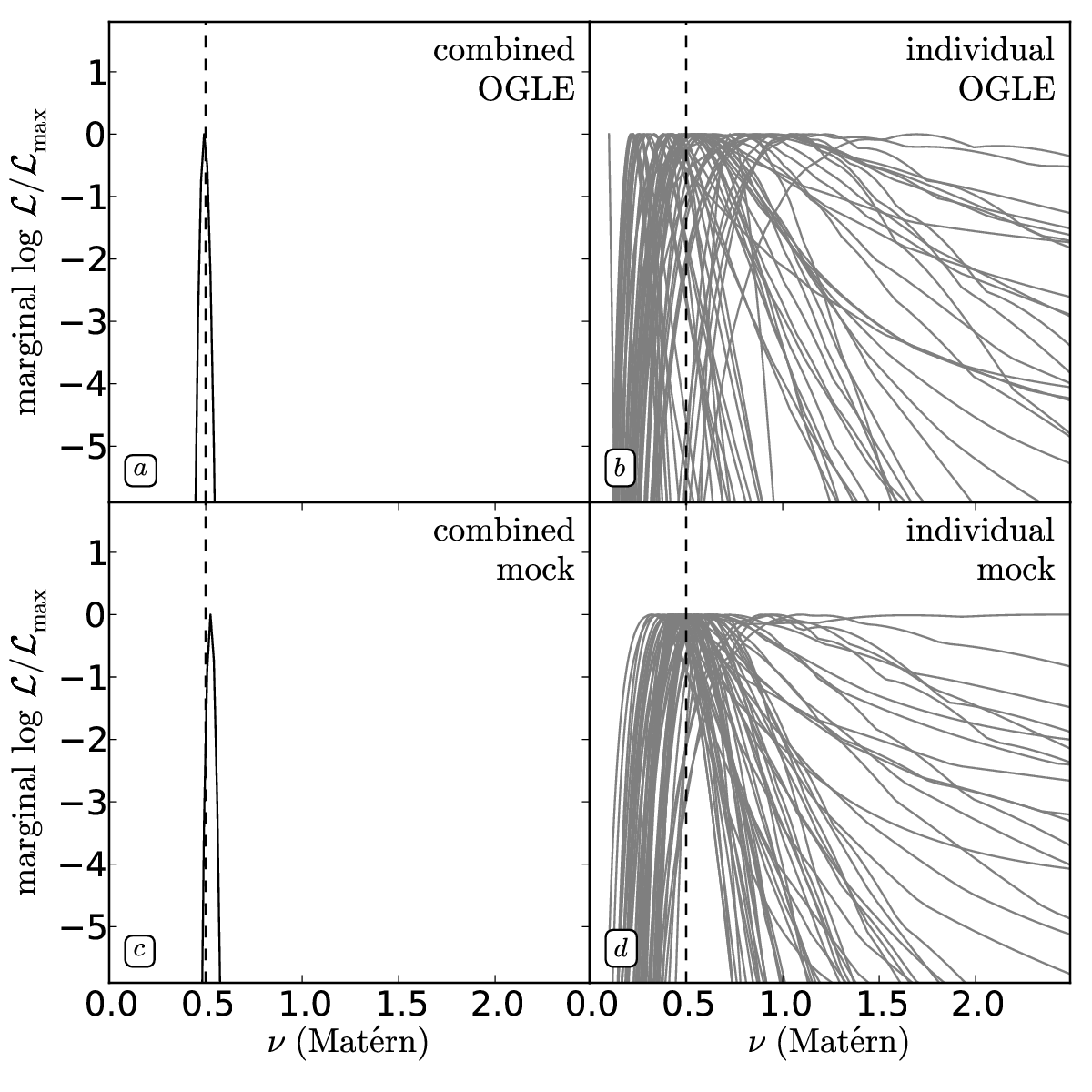} 
\caption{As in
Figure~\ref{fig:pe} but for the \ma~model. Here $\nu=0.5$ corresponds to the DRW
model.} 
\label{fig:ma} 
\end{figure}

We first fit each quasar light curve with the DRW model to obtain Maximum
Likelihood~(ML) estimates for $\sigma$ and $\tau$, which are then used to
generate a mock light curve that is fully consistent with the DRW model while
having exactly the same time sampling and photometric uncertainties as the true
light curve. We generate the DRW mock light curves as random GP realizations
using the Cholesky decomposition method described in~\cite{zu2011}, which is a
generic method for all covariance function models. DRW mock light
curves can also be easily generated as CAR(1) processes~(see
Footnote~\ref{note:simu},~\cite{kelly2009}). Finally, we add a Gaussian deviate
normalized by the estimated photometric error of each data point to
incorporate the measurement uncertainties of the data into the mock light
curves. We use these mock DRW light curves as a comparison sample when
evaluating the evidence for any additional parameters. 

Next, we fit both the data and the mock light curves using each
of the four covariance functions.  Since we only care about
the differences in the third parameter, we start with a Maximum
Likelihood analysis based on $\mathcal{L}(\text{$[\gamma, \nu, \tcut,
\alpha]$})\equiv\mathcal{L}(\tau_{max}, \sigma_{max}, \text{$[\gamma,
\nu, \tcut, \alpha]$})$ where $\tau_{max}$ and $\sigma_{max}$ maximize
the likelihood at fixed 3rd parameter for each light curve. In addition to
considering the models of the individual light curves, which we will refer
to as the individual models, we also want to derive the mean and dispersion
of the third parameter of each model for the ensemble light curves. However,
since the individual likelihoods are mostly asymmetric with non--Gaussian
tails, it is inappropriate to calculate the weighted mean directly from
the best--fit third parameters of individual models. To fully take into
account the shape of individual likelihoods, we multiply~(add) all the
individual (log)likelihood functions together to obtain a ``combined''
likelihood function of the third parameter.  By comparing the ``combined''
and individual likelihood functions for the data and mock light curves, we
can then assess how significant the best--fit third parameter deviates from
or agrees with the DRW prediction in each of the four modifications. We will
focus on this as a problem of parameter estimation rather than model testing,
although we will discuss model testing.

\section{Results}
\label{sec:result}

\begin{figure}[t] 
\epsscale{1.00} 
\plotone{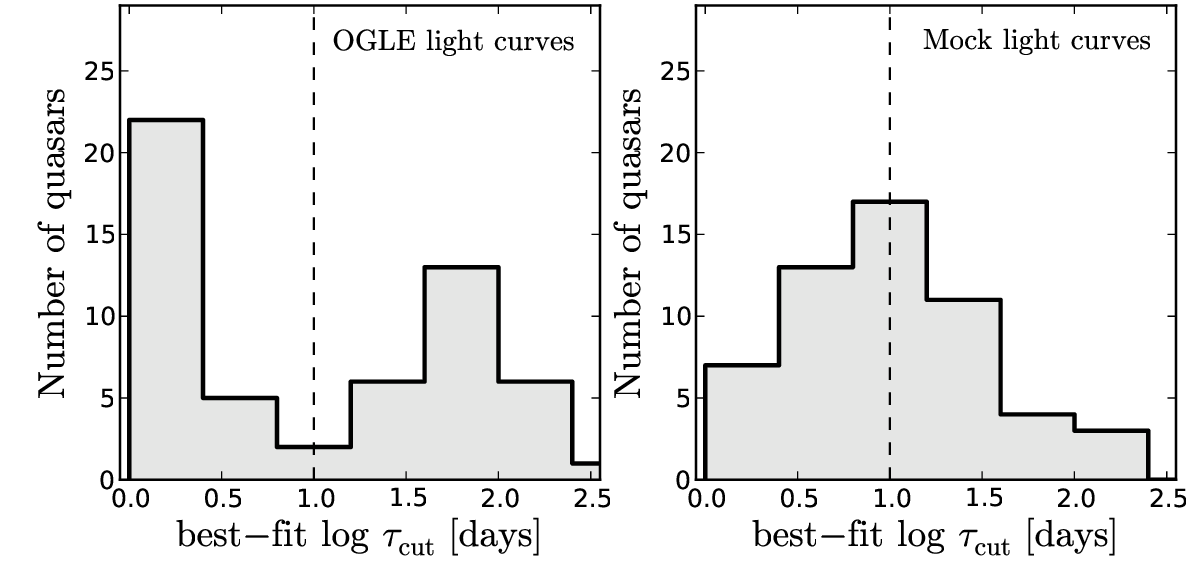} 
\caption{Distribution of quasars as a
function of the best--fit $\tcut$ for the data~(left panel) and mock~(right
panel) light curves using the \ke~model. The vertical dashed line in each panel
at $10$ days indicates the scale below which no variability
signal can be resolved for the lack of sampling the \ke~covariance function below $\tcut$.}
\label{fig:keplertest}
\end{figure}

The results of our model depend on how well the DIA photometric errors
of the light curves in our sample are determined. Assuming the DRW is
an adequate model for each individual light curve,\footnote{The DRW
model has one fewer degree of freedom than the other models, so the inferred
error statistics should be more conservative.}
we can estimate the fractional errors in the photometric error estimates
from the $\chi^2/{dof}$ distribution of the DRW fits to the 55 light
curves. For all these objects, the overall goodness of fits to the light
curves with the DRW model are reasonable~($0.88<\chi^2/{dof}<1.53$),
with an average $\left<\chi^2/{dof}\right>\simeq 1.09$ and
dispersion $\sigma_{\chi^2/{dof}}=0.13$, while we would expect
$\left<\chi^2/{dof}\right>=1$ and $\sigma_{\chi^2/{dof}}=\sqrt{2/{dof}}\simeq
0.05$. These differences could be explained by fractional shifts $\sigma_m =
\sigma_m^{true}(1+e)$ between the true uncertainties $\sigma_m^{true}$ and the
nominal error bar $\sigma_m$ where the mean fractional error is $\left<e\right>
=-0.04$ with a dispersion of $\sigma_e=0.06$. This simple assumption on the
OGLE photometric uncertainties will help us understand the scatter in any
additional model parameter between quasars, but to cover all the possibilities,
we discuss the implications of our results in terms of two scenarios: (1)
that the OGLE photometric uncertainties are correct to high accuracy~($\sim
10\%$) and lack significant non--Gaussian tails, (2) that the uncertainties are
slightly underestimated and/or have some non--Gaussian tails. We hereby refer to
the two scenarios as case I and case II, respectively.

We simply discuss each of the four models in turn, starting with the PE model
results shown in Figure~\ref{fig:pe}. The panels on the left show that the
combined results, where we merge the likelihood functions of all 55 quasars,
agree with the DRW model remarkably well. The joint marginal
likelihood function for $\gamma$ strongly peaks near $\gamma=1.0$ with
$\left<\gamma_\mathrm{data}\right> = 0.98 \pm 0.04$ using $\Delta
\ln(\mathcal{L}/\mathcal{L}_\mathrm{max}) = -2.0$ for the error estimate~(which
would correspond to $2\sigma$ for Gaussian uncertainties). This is generally
consistent with the result for the mock light curves of
$\left<\gamma_\mathrm{mock}\right> = 1.04 \pm 0.03$.  The individual
likelihoods, shown in the right panels of Figure~\ref{fig:pe}, seem to show a
larger scatter in the data~($\sigma_{\gamma, \mathrm{data}}=0.37$) than in the
mock data~($\sigma_{\gamma, \mathrm{mock}}=0.19$). While essentially all the
light curves are consistent with the DRW at better than
$3\sigma$~($\left<\Delta \ln(\mathcal{L}/\mathcal{L}_\mathrm{max}) =
-4.2\right>$), there is one extreme outlier~(seen as the spike at
$\gamma=0.14$).  While there is nothing obviously odd in this object's light
curve, it is very close to our noise selection limit and a source will be
biased towards low $\gamma$~(closer to white noise) if the true measurement
errors are larger than estimated~(see below). This object is also an outlier in
the other models. Nonetheless, excluding or including it has a
negligible effect on the combined likelihood functions.

Before going to other models, we want to interpret the larger spread
we observe in the best--fit $\gamma$ for the individual data light
curves~(panel $b$ of Figure~\ref{fig:pe}). The difference of the
dispersions between the individual best--fits in the data and mock light
curves is $\sigma_{\gamma}=(\sigma_{\gamma, \mathrm{data}}^2-\sigma_{\gamma,
\mathrm{mock}})^{1/2}=0.31$~($0.29$ if we drop the one extreme outlier). This
can be interpreted either as evidence for intrinsic scatter if case I
is correct, or as evidence for problems in the photometric uncertainties
$\sigma_m$~(case II). As an experiment, we re-fit the PE model to the mock
light curves after resetting $\sigma_m$ to be
$20\%$ larger~($e=+0.2$) or smaller~($e=-0.2$) than the actual uncertainties
used to generate the mock light curves. The results of the experiment are shown
in Figure~\ref{fig:errtest}. When the noise is over-estimated, true variability
power on short time scales is instead interpreted as noise, so the models shift
to higher $\gamma$ for stronger correlations on short time scales. When the
noise is under-estimated, noise on short time scales is interpreted as signal,
so the models shift to lower $\gamma$ for weaker correlations. As expected, in
Figure~\ref{fig:errtest}, over-estimating the errors biases the estimates of
$\gamma$ to be high, with $\left<\gamma_\mathrm{mock}^{+}\right> = 1.30 \pm
0.04$~(panel a of Figure~\ref{fig:errtest}), and vice versa when
under-estimating errors, $\left<\gamma_\mathrm{mock}^{-}\right> = 0.62 \pm
0.01$~(panel c of Figure~\ref{fig:errtest}). Given that a $20\%$ error in
$\sigma_m$ causes a $0.3$ bias in the estimate of $\gamma$, so that
$\gamma\simeq1+1.5e$, then the excess variance in the estimates of $\gamma$ for
the data could be explained by making $\sigma_e=0.20$, which is significantly
larger than the $\sigma_e=0.06$ suggested by the $\chi^2/{dof}$ distribution.
Also note that the estimate of $\left<e\right>$ from the $\chi^2/{dof}$
distribution would tend to shift the estimates of $\gamma$ for a DRW model to
$\gamma = 1 + 1.5\left<e\right>\simeq0.93$, so systematic uncertainties in the
photometric errors produce uncertainties comparable to the statistical
uncertainties in $\left<\gamma\right>$. If we use $\sigma_e=0.06$ from the
$\chi^2/{dof}$ distribution as an estimate of the scatter in the fractional
uncertainties in $\sigma_m$~(case I), then we appear to be left with
$\sigma_{\gamma}^{in}=0.28$ of intrinsic scatter between quasars. If case II is
correct~(i.e., $\sigma_e\sim20\%$ or strong non--Gaussian tails in $\sigma_m$),
the observed mean and spread of $\gamma$ are consistent with DRW model with
little quasar--to--quasar variation.

For two other models~(MA and PA, discussed further below), we also see
evidence for larger spread in the best--fit third parameter in the data than
in the mock data.  We have done similar experiments of scaling the photometric
uncertainties for each model and the results stay consistent with our findings
for the PE model and the two possible interpretations. To avoid unnecessary
repetition, we will only report the intrinsic scatters inferred from each
experiment for the two models.

\begin{figure}[t] 
\epsscale{1.00} 
\plotone{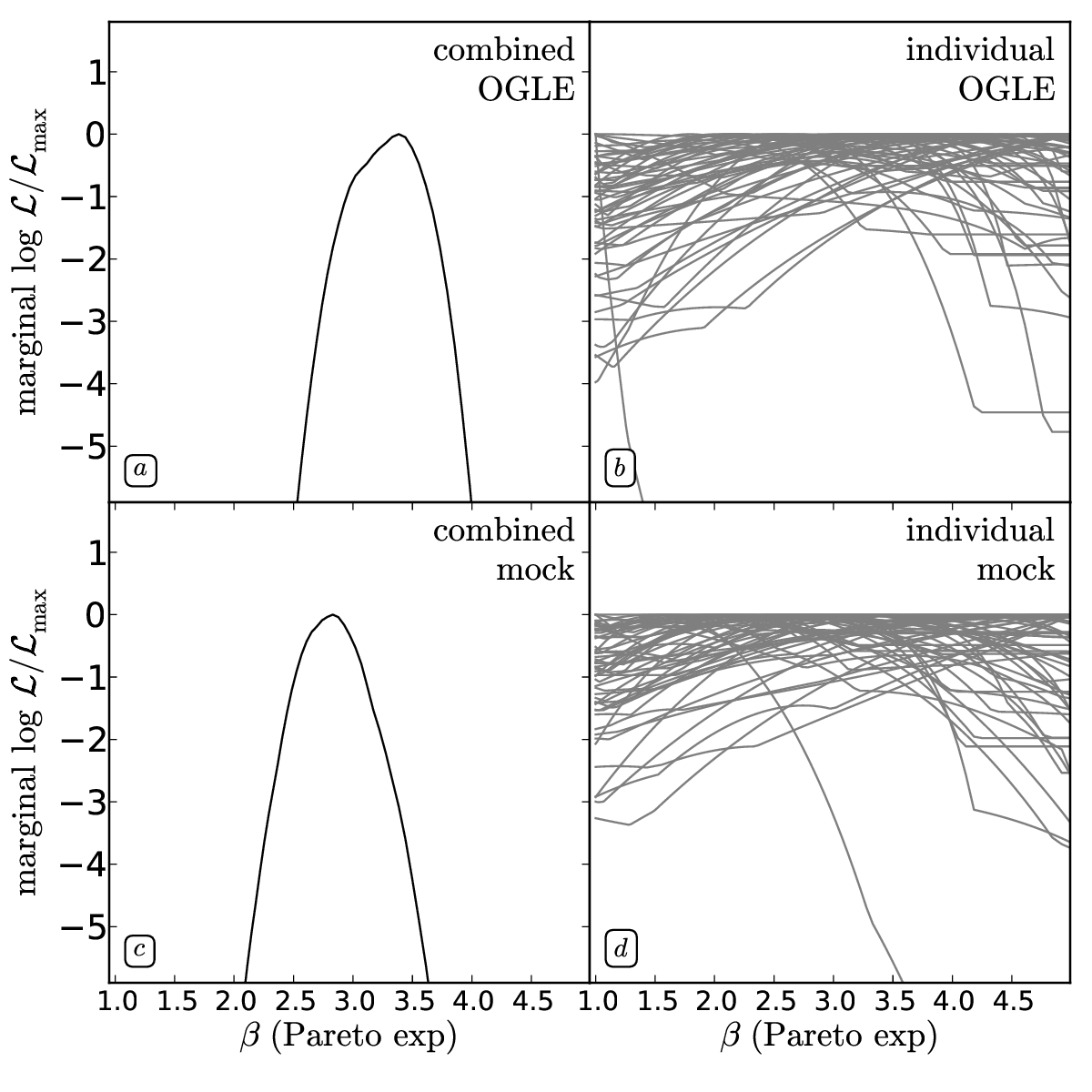}
\caption{As in Figure~\ref{fig:pe} but for the \pa~model. No value of $\beta$
corresponds to the DRW model.}
\label{fig:pa}
\end{figure}

Figure~\ref{fig:ma} shows the results for the MA model.  The
likelihood function for the data is again strongly peaked at the value that
corresponds to the DRW model, $\nu=0.5$, with $\left<\nu_\mathrm{data}\right> =
0.492 \pm 0.016$, as compared to $\left<\nu_\mathrm{mock}\right> = 0.525 \pm
0.017$. The distribution of the individual results, is again broader for the
real data than for the mock data, suggesting an excess scatter of
$\sigma_{\nu}=0.08$ in $\nu$ if there are no other systematic uncertainties. In
this case, the sensitivity to the uncertainties in the photometric errors is
$\nu = 0.5 + 1.1e$, suggesting an intrinsic scatter of $\sigma_{\nu}^{in}=0.05$
in case I and broad consistency with DRW model in case II.
 
On even shorter time scales, the histograms in Figure~\ref{fig:keplertest} show
the distributions of the best--fit $\tcut$ of the KE model for the data~(left
panel) and the mock~(right panel) light curves. For the data, there is a clearly
bimodal distribution of $\tcut$ demarcated by a gap around $10$ days that
corresponds to 5 times the typical cadence~($2$ days). The first peak below 10
days comprises objects that are either consistent with the DRW model~($\tcut=0$)
or have a $\tcut$ that is unresolved by the OGLE sampling~($\leqslant 4$
discrete values of $\dt$ below $\tcut$ could be computed for $\cov_\mathrm{KE}$
in Equation~\ref{eqn:covke}). The second peak at $30-100$ days is well beyond
the cadence scale, indicating another population of objects~(23 out of 55) that
favor a cutoff time scale beyond a month. For the mock sample, there is only a
single peak below $10$ days and a steady decrease towards longer $\tcut$, with
only 11 out of 55 objects having $\tcut$ larger than a month.  Unfortunately,
the estimates of $\tcut$ are very susceptible to errors in the estimated
photometric errors, so the second peak in the $\tcut$ distribution of the data
light curves could also be an artifact of the scatter in $e$. Thus a cutoff time
scale $\tcut\sim$ 1--3 months may be marginally detected in half of the quasars,
but the cadence and photometric uncertainties of the data are not optimal for
probing this regime.



Since the DRW model is a restricted form of the \pe, \ma, and \ke~models, with
$\gamma=1$, $\nu=0.5$, and $\tcut=0$, respectively, we could use a standard
$\mathcal{F}$--test to evaluate whether the DRW model is preferred.  We must simply
treat the matrix determinant in the likelihood~(Equation~\ref{eqn:likelihood})
as an additional contribution to the $\chi^2$, so $\chi^2\equiv-2\ln\mathcal{L}$.
If we adopt a probability threshold of $p>5\%$, 50~(PE), 51~(MA), and 53~(KE)
out of 55 mock light curves are consistent with the null hypothesis that
no additional parameter is needed. For the real light curves, 34, 33, and
53 of the 55 light curves have $p>0.05$ for the PE, MA, and KE models,
respectively.  Thus we again find that the light curves are generally
consistent with the DRW model but with some scatter about it. The problem,
however, is that the $\mathcal{F}$--test is not truely applicable to the
likelihood computed in Equation~\ref{eqn:likelihood}. A more appropriate
approach is to compare the three models using a Bayesian framework. In
this approach, the addition of parameters is penalized by modifying the
likelihood. Two standards are the Bayesian Information criteria~(BIC;
$\log\mathrm{L} - 0.5 k \log n$ where $k$ and $n$ are the
numbers of model parameters and data points, respectively) and the Akaike
Information criteria~(AIC; $\log\mathrm{L} - k$), The two
criteria are very similar, except that the BIC more heavily penalizes the
addition of parameters. The DRW model is overwhelmingly favored by the more
``conservative'' BIC, while for the more ``liberal'' AIC, the probability
ratios are $P_\mathrm{DRW}:P_\mathrm{PE}:P_\mathrm{MA} = 2:1:1$.

Finally, the light curves do not constrain changes in the structure of
covariance function on longer timescales. The individual likelihood functions
for $\alpha$ in the \pa~model are mostly flat, as shown in Figure~\ref{fig:pa}.
The combined constraint from the data favors a slightly higher value of
$\left<\alpha_\mathrm{data}\right>=3.38 \pm 0.45$ than the mock light curves,
$\left<\alpha_\mathrm{mock}\right>=2.83\pm 0.42$, but the two peaks are
mutually consistent. The uncertainties are, however, too large to justify
additional models for the long time scale behaviors.

\section{Discussion}
\label{sec:dis}

We considered four families of covariance functions to test whether there
is any evidence that quasar optical variability differs from a DRW model
on time scales of weeks to years. Our conclusion is that deviations may
exist, but they are small enough that we are reluctant to draw a firm
conclusion. The significance of the deviations is very sensitive to the
accurary of the photometric errors, and different tests lead to different
statistical conclusions. $\mathcal{F}$--tests, which are not really correct
given our likelihood functions, imply the differences are real, while
Bayesian methods, which are more applicable, imply the modifications are
unnecessary or only equally probable as compared to the default DRW model,
depending on the information criterion used to evaluate the question.  Thus,
we will discuss the results in terms of both possibilities. More specifically,
\begin{itemize}
\item On the time scales best sampled by the light curves~(from months to
years), the typical light curve is generally well-described by the DRW model,
potentially with some intrinsic scatter in the true structure of the power spectrum if
the photometric error estimates are correct to better than $10\%$. The observed
scatter could also be extrinsic if the uncertainties in photometric error
estimates are larger than $10\%$ or the error distributions have non--Gaussian tails.
\item On very short time scales, there are hints of a characteristic cutoff time
scale $\tcut$~(1--3 months) for $\sim 50\%$ of the quasars, below which the
correlations become stronger than predicted by the DRW model. 
This is consistent with the Kepler results of~\cite{mushotzky2011}, but the
OGLE data are clearly not competitive with {\it Kepler} in this regime.
\item On very long time scales, the light curves are consistent with the DRW
model, but the precision of the test is limited by the length~($7$ years) of
the light curves.  
\end{itemize}

Under the assumption that the photometric error estimates are broadly correct to
with $10\%$, the apparent scatter about the DRW model may well be evidence for multiple
stochastic processes rather than a single DRW. For example, the \pe~covariance
function can be viewed as the mixture of a continuous sum of independent
DRW models with characteristic timescale $\tau_i$,
\begin{equation}
\sigma^2\exp \left(-\left|\frac{\dt}{\tau}\right|^\gamma\right) = \int_0^{+\infty} P(s,
\gamma)\sigma^2(s\tau)
\exp\left(-\left|\frac{\dt}{s\tau}\right|\right)ds,
\end{equation}
where $s\equiv \tau_i/\tau$, $\sigma^2(s\tau)$ is the amplitude of process $i$,
and $P(s, \gamma)$ is the probability density function for the process on time
scales $\tau$. The combination of $P(s, \gamma)\sigma^2(s\tau)$ is then the
inverse Laplace transform of the powered exponential~\citep{johnston2006}. For
$\gamma=1$~(DRW), $P(s, \gamma)$ is a Dirac $\delta$ function at $s=1$, so there
is only one process operating at one timescale $\tau$. Adding several weaker
processes with different time scales would then appear as a scatter about
$\gamma=1$, but would be difficult to distinguish from a single DRW model. This
is similar in spirit with the work of ~\cite{kelly2011}, where they modeled
quasar optical variability as a linear mixture of different DRW processes and
found it generally provides no better fit than a single DRW model. In the toy
model adopted by ~\cite{dexter2011}, they matched the observed DRW variability
by exciting many independent fluctuating zones with exactly the same $\tau=200$
days in the accretion disc, and then argued that the resulting change in the
effective temperature profile of the disk would explain the observed
discrepancies between thin disc sizes inferred from gravitational microlensing
of lensed quasars and their optical luminosities~\citep{morgan2010}.

It would be useful to have occasional campaigns in which
OGLE sampled shorter time scales with longer integration times to fill in these
time baselines with higher precision data. On long time scales, the OGLE-IV
project is already extending the light curves, so it will become increasingly
possible to explore the long time scale behavior of individual quasars.
As the behavior on long time scales~(decades) becomes better constrained,
it will also be easier to constrain the short time scales~(months to years)
because there will be less freedom due to covariances between the parameters.
The continuing spectroscopic surveys by~\cite{kozlowski2012} will also
greatly increase the number of confirmed quasars with OGLE light curves.
DES~\citep{the_dark_energy_survey_collaboration2005}, LSST~\citep{ivezic2008}
and Pan-STARRS~\citep{kaiser2002} will carry out similar extensions for the
SDSS Stripe 82 quasars but generally at lower cadence (albeit for larger
numbers of objects). Given a larger sample it wll also be possible to search
for correlations of the apparent deviations from the DRW model with other
physical characteristics of the quasars (luminosity, wavelength, etc.).

\section*{Acknowledgements}
C.S.K is supported by the NSF grant AST-1009756. OGLE is supported by the
European Research Council under the European Community's Seventh Framework
Programme (FP7/2007-2013), ERC grant agreement no. 246678. S.K. is supported by
the Polish Ministry of Science and Higher Education (MNiSW) through the
``Iuventus Plus'' program, award number IP2010 020470.


%
\end{document}